\documentclass[12pt,preprint]{aastex}

\shorttitle{X-ray Polarimetry}
\shortauthors{Strohmayer et al.}

\begin{document}


\title{X-ray Spectro-polarimetry with Photoelectric Polarimeters}


\author{T. E. Strohmayer}
\affil{X-ray Astrophysics Lab, Astrophysics Science Division, NASA's Goddard 
Space Flight Center, Greenbelt, MD 20771}

%





\begin{abstract}

We derive a generalization of forward fitting for X-ray spectroscopy to
include linear polarization of X-ray sources, appropriate for the
anticipated next generation of space-based photoelectric polarimeters.
We show that the inclusion of polarization sensitivity requires joint
fitting to three observed spectra, one for each of the Stoke's
parameters, $I(E)$, $U(E)$, and $Q(E)$. The equations for Stoke's
$I(E)$ (the total intensity spectrum) are identical to the familiar
case with no polarization sensitivity, and for which the
model-predicted spectrum is obtained by a convolution of the source
spectrum, $F(E')$, with the familiar energy response function,
$\epsilon(E') R(E', E)$, where $\epsilon(E')$ and $R(E', E)$ are the
effective area and energy redistribution matrix, respectively. In
addition to the energy spectrum, the two new relations for $U(E)$ and
$Q(E)$ include the source polarization fraction and position angle
versus energy, $a(E')$, and $\psi'_0(E')$, respectively, and the
model-predicted spectra for these relations are obtained by a
convolution with the ``modulated'' energy response function,
$\mu(E')\epsilon(E')R(E, E')$, where $\mu(E')$ is the energy-dependent
modulation fraction that quantifies a polarimeter's angular response
to $100\%$ polarized radiation. We present results of simulations with
response parameters appropriate for the proposed {\it PRAXyS} Small
Explorer observatory to illustrate the procedures and methods, and we
discuss some aspects of photoelectric polarimeters with relevance to
understanding their calibration and operation.

\end{abstract}





\section{Introduction}

In comparison with imaging, timing and spectroscopic measurements,
polarization remains the ``missing piece of the puzzle'' of
observational X-ray astrophysics. To date, the only measurement in the
2 - 10 keV band is still the $\approx 20\%$ polarization fraction
inferred for the Crab nebula (Weisskopf 1978). The dearth of
additional detections has largely been because of a lack of
instruments sensitive enough to make such observations.  However, with
the advent of micro-pattern gas detectors which can directly leverage
the photoelectric effect to infer linear polarization (Costa et
al. 2001; Black et al. 2004, 2010), it is likely that this situation
will change in the not-too-distant future. Indeed, a number of X-ray
polarimetry mission concepts have been proposed in the last few years.
Among these are several photoelectric effect polarimeters sensitive in
the 2 - 10 keV band, including the Polarimeter for Relativistic
Astrophysical X-ray Sources ({\it PRAXyS}) Small Explorer (Hill et
al. 2014; Jahoda et al. 2016), and the Imaging X-ray Polarimetry
Explorer ({\it IXPE}, Weisskopf et al. 2016). Also, the proposed {\it
  PolSTAR} experiment (Krawczynski et al.  2015) would pair a hard
X-ray mirror similar to that flown on {\it NuSTAR} with a passive
scattering element to provide broad band X-ray polarimetry from
$\approx 3 - 50$ keV.  Above 10 keV Compton scattering becomes
competitive with the photoelectric effect, and several instruments
have recently been developed to exploit this to enable polarimetry in
the hard X-ray band.  Among these are the balloon-borne payloads {\it
  X-Calibur} (Beilicke et al. 2014) and {\it PoGOLite} (Chauvin et
al. 2016). Additionally, ESA has recently selected the X-ray Imaging
Polarimetry Explorer ({\it XIPE}, Soffitta et al. 2016) for study and
possible implementation as ESA's M4 medium class mission.  Finally, we
note that shortly after submission of this paper, NASA selected {\it
  IXPE} for implementation in the 2020 timeframe as a Small Explorer
mission (SMEX).

In anticipation of the further opening of the polarization window in
the X-ray band it is timely to explore the question how one can
properly generalize X-ray spectroscopic observations to include linear
polarization of X-ray sources.  That is, what is the additional
computational ``machinery'' required to do X-ray spectro-polarimetry
from space observatories.  In this paper we outline in some detail how
to infer the physical properties of sources which include linear
polarization properties with space-based photoelectric
polarimeters. This includes a generalization of the standard methods
of X-ray spectral ``forward fitting'' to include polarization
properties, as well as a discussion of the detector calibration
information that is needed.  

We also discuss some aspects of photo-electric polarimeters relevant
to understanding their calibration and operation, and we present the
results of simulations that illustrate the procedures and methods
using spectro-polarimetric capabilities appropriate for the proposed
{\it PRAXyS} Small Explorer observatory.  We note that the methods
described here should also prove applicable to data expected from
NASA's recently selected {\it IXPE} small explorer observatory, as
well as other future X-ray polarimetry missions.

\section{Background}

For the simpler, well-known case where X-ray detectors have no
polarization sensitivity, an observation of an astrophysical source,
at least in the context of spectroscopy, can be characterized by a
physical input (intrinsic) source spectrum, and the detector's energy
response function, which quantifies the rate at which photons of
intrinsic energy $E'$ are observed in detector energy channel $E$. The
energy response function is often broken up into two components, the
energy redistribution matrix, $R(E',E)$, which is defined as the
probability that a photon of energy $E'$ is detected in channel $E$,
and the effective area, $\epsilon(E')$, which quantifies the
detector's collecting area as a function of energy.  The count rate
spectrum predicted to be observed in a detector is then the
convolution,
\begin{equation} 
O(E) = \int_{E'} F(E') \epsilon(E') R(E',E) dE' \; ,
\end{equation}
where $F(E')$ is the intrinsic source spectrum, that is, the number of
source photons with energies between $E'$ and $E' + dE'$. The
standard, ``forward fitting'' procedure for doing X-ray spectroscopy
is then to introduce some model parameterization for $F(E')$, use that
model and the response functions to generate predicted detector
channel spectra, and then constrain the model's parameters by
comparing these to the actual observed spectra using a statistical
procedure such as $\chi^2$ fitting (see, Lampton, Margon \& Bowyer
1976). For example, the XSPEC software package is a commonly used tool
in the X-ray astrophysics community to implement this procedure
(Arnaud 1996).  

\section{Polarimetry}

The generalization to spectro-polarimetry adds an additional
observable, the sky angle, $\psi$, of the polarization vector of each
detected photon. A convenient way to define the sky angle is to
construct a coordinate system on the sky with $x$ and $y$ axes defined
as local North and East, respectively. The polarization angle is then
taken as the azimuthal angle increasing in the counter-clockwise
direction when looking at the sky. The $z$-axis points from the source
to the observer, forming a right-handed coordinate system.  It is
standard to define local North as having an angle of zero, and then an
angle pointing East is $\pi/2$.  For linear polarimetry a complete
description can be given with this angle defined on the range from $0$
to $\pi$. We note that this description is consistent with the IAU
convention for polarimetry (IAU 1974; see also Hamaker \& Bregman
1996).

Whereas with no polarization sensitivity the detector response matrix
is a two dimensional function, for polarimetry it becomes a four
dimensional function, $X(E', E, \psi', \psi)$, that describes the rate
at which photons with intrinsic energies $E'$ and intrinsic
polarization angles $\psi'$ are observed by a detector with energy $E$
and a 100\% polarization along the angle $\psi$.  A good polarimeter
will be one where the observed angular distribution in $\psi$ of the
counts in any detector energy channel (or range of channels) is
well-peaked around the intrinsic angle $\psi'$, the phrase ``more
strongly modulated'' around $\psi'$ also comes to mind.

We consider a general X-ray source as an incoherent ensemble of
photons with wave-vectors $\vec k = \hat z$. Such an ensemble can
always be represented as a superposition of 100\% polarized photons.
In this more general case the observed ``spectrum'' is a function of
both energy and sky angle. For a given energy channel (or range of
channels) the distribution of observed sky angles is commonly referred
to as a modulation curve, which can be written in a general way as,
\begin{equation}
O(E, \psi) = \int_{E'} \int_{\psi'} H(E', \psi') X(E', E, \psi', \psi)
dE' d\psi' \;,
\end{equation}
where $H(E', \psi')$, is now a more general intrinsic source spectrum
that describes the number of source photons with energy $E'$ between
$E'$ and $E' + dE'$, and with polarization angles between $\psi'$ and
$\psi'+d\psi'$.  We will introduce several general source descriptions
specific to linear polarization, including a Stokes parameter
description, shortly.  

In practice, a detector system will produce a polarization angle
measurement in a coordinate system defined by the detector's geometry,
mode of operation, and orientation. We call this the detector frame,
and we refer to intrinsic and measured polarization angles in this
frame as $\phi'$ and $\phi$, respectively. The orientation of the
detector with respect to the sky coordinate system defines a mapping
which relates the sky angle and detector frame angles. We define this
mapping as $\rho(t)$, such that $\psi' = \phi' + \rho(t)$, and $\psi =
\phi + \rho(t)$.  If the detector orientation is fixed in time, then
the mapping is simply a constant offset, $\rho_0$. If however the
detector is rotated at a constant rate about the line of sight (say,
by the planned rolling of a spacecraft around the line of sight), then
the mapping is a simple linear function of time, $\rho(t) = \Omega t$,
where $\Omega$ is the angular roll rate of the spacecraft.

Now, let's begin by exploring an {\it ansatz} where the detector
response function, $X(E', E, \phi', \phi)$, can be factored into two
terms, the standard energy response function, and a new term, $V(E',
\phi', \phi)$, that describes the angular response of the
polarimeter. This term effectively describes how well a photon's
intrinsic polarization angle $\phi'$ can be measured.  With this
description the full response function can be written,
\begin{equation}
X(E', E, \phi', \phi) = \epsilon(E')R(E',E) V(E',\phi', \phi) \; .
\end{equation}
Specifically, the angular term, $V(E', \phi', \phi)$, describes the
probability for a photon of intrinsic energy $E'$ and intrinsic
polarization angle $\phi'$ to be observed with a polarization angle
$\phi$.  

We note that there could be additional ways of considering how to
construct $X(E', E, \phi', \phi)$, however, we argue that the above is
reasonable for several reasons. First, It is known that grazing
incidence X-ray optics essentially do not alter the polarization
properties of incident photons at levels relevant for this discussion
(see Almeida \& Pillet 1993; Chipman et al. 1992, 1993; Hill et
al. 2016).  Second, the dominant photoelectric cross-section in the
relevant energy band (ionization of K shell electrons in an $s$ state)
is not dependent on the polarization angle of the photon, and while
some materials can display linear dichroism, a variation in the
absorption strength with polarization direction, this is generally a
small effect and only confined to narrow bands around some absorption
edges (see for example, Collins 1997; Bannister et al. 2006). Thus, to
good approximation the relevant physics associated with $\epsilon(E')$
and $R(E',E)$ is largely independent of the photon polarization
direction. Finally, one will ultimately determine the response
function in an empirical fashion, by calibrating a detailed physical
model of the detector system against actual measurements.  For
example, one can carry out laboratory measurements whereby $100\%$
polarized beams of photons of known energy and polarization angle
illuminate the detector.  The observed modulation curves for different
input photon energies can then be used as a direct estimator of the
angular distribution $V(E', \phi', \phi)$, and one can explore
empirically the extent to which equation (3) is realized in practice.

For photoelectric polarimeters, which infer the angular direction of
an ejected photoelectron, the angular response function is generally
of the form, $V(E', \phi', \phi) = v(E', (\phi - \phi')) = C_0 (A(E')
+ B(E')\cos^2 (\phi - \phi'))$, where $C_0$ is an overall
normalization factor (see, for example, Costa et al. 2001).  From this
expression one can define the energy-dependent modulation fraction,
$\mu(E') \equiv (V_{max} - V_{min} )/(V_{max} + V_{min} ) =
B(E')/(2A(E') + B(E'))$ (Costa et al. 2001; Strohmayer \& Kallman
2014). Alternatively, one can also write this response using an
equivalent Stokes formalism as, $V(E',\phi', \phi) = C_0(i(E') +
u(E')\sin(2\phi) + q(E')\cos(2\phi))$, where now the energy-dependent
modulation fraction is $\mu(E') = (u^2(E') + q^2(E'))^{1/2} / i(E')$,
and $\phi' = 1/2\tan^{-1} (u(E')/q(E'))$. A proper angular response
function must be normalized such that the integrated probability gives
unity.  It is straightforward to show that for the function $V(E',
\phi', \phi) = C_0 (A(E') + B(E')\cos^2 (\phi - \phi'))$, and with
$\phi$ ranging from $0$ to $\pi$, $C_0 =( \pi (A(E') + B(E')/2
))^{-1}$.

\subsection{Source Descriptions}

First, we emphasize that photoelectric polarimeters are only sensitive
to linear polarization, and not, circularly polarized radiation. That
is, a purely circularly polarized X-ray beam would appear
``unpolarized,'' to such a detector, in the sense that the observed
angular distribution of polarization position angles would be
identical to that produced by a completely unpolarized beam. Thus,
when we refer to an unpolarized spectrum or flux, this technically
also includes any circularly polarized component of the source flux.
For the case of linear polarization the spectro-polarimetric
properties of X-ray sources can be described in several equivalent
ways.  A convenient description uses the so-called Stokes parameter
decomposition. Here one can define the source spectrum as,
\begin{equation}
H(E', \psi') = F(E') + W(E') + Z(E') \; ,
\end{equation}
where $W(E')$ and $Z(E')$ describe the linear polarization properties
of the source, and $F(E')$ is the so-called total intensity energy
spectrum.  The fractional polarization amplitude is then given by,
$a(E') = (W^2 (E') + Z^2(E'))^{1/2} / F(E')$, and the source
polarization position angle is $\psi'_0(E') = 1/2 \tan^{-1}
(W(E')/Z(E'))$. An equivalent description can be given using the
un-polarized, $h(E')$, and polarized, $g(E')$, spectra,
\begin{equation}
H(E', \psi') = h(E') + g(E') \; ,
\end{equation}
and one must also define the energy-dependent polarization position
angle, $\psi'_0(E')$. In this case the polarization amplitude is
$a(E') = g(E') / (h(E') + g(E'))$. A convenient feature of this
prescription is that each spectral component has a simple angular
dependence. By definition, the un-polarized spectrum, $h(E')$, is
uniformly distributed with intrinsic polarization angles $\psi'$
ranging from $0$ to $\pi$, and the polarized spectrum, $g(E')$,
only has intrinsic angles $\psi'(E') = \psi'_0(E')$. These two
descriptions being equivalent, it is straightforward to show that the
polarized spectrum is $g(E') = ( W^2(E') + Z^2(E'))^{1/2}$, and the
un-polarized spectrum is $h(E') = F(E') - ( W^2(E') + Z^2(E'))^{1/2}$.

When generating physical source models researchers may find it more
convenient to compute the total spectrum, $F(E')$, as well as the
energy-dependent polarization amplitude, $a(E')$, and position angle,
$\psi'_0(E')$.  In such a case it is then straightforward to determine
the Stokes spectra, $W(E')$ and $Z(E')$ from the above
equations. Doing this one finds,
\begin{equation}
W(E') = \frac{F(E')a(E')\tan(2\psi'_0(E'))}{(1 + \tan^2 (2\psi'_0(E'))
  )^{1/2} } = F(E')a(E')\sin(2\psi'_0(E')) = g(E')\sin(2\psi'_0(E'))
\end{equation}
and 
\begin{equation}
Z(E') = \frac{F(E')a(E')}{(1 + \tan^2 (2\psi'_0(E')) )^{1/2} } =
F(E')a(E')\cos(2\psi'_0(E')) = g(E')\cos(2\psi'_0(E')) \; .
\end{equation}

\subsection{X-ray Spectro-polarimetry}

We can now use equation (2) above to determine the observed spectrum
for a particular source description and detector response
functions. For the source description we will use $H(E',\psi') = h(E')
+ g(E')\delta(\psi' - \psi'_0(E'))$ since each component has a
well-defined angular distribution, which will simplify the
integrations over $\psi'$. Here, the Dirac delta function restricts
the intrinsic polarization angles, $\psi'$, for the polarized spectrum
to the position angle for the source, $\psi'_0(E')$. We will also use
$V(E', \phi', \phi) = C_0 ( A(E') + B(E')\cos^2 (\phi - \phi'))$. Note
that with this form for $V(E', \phi', \phi)$, we have that $\cos^2
(\phi - \phi') = \cos^2 ( (\psi - \rho(t)) - (\psi' - \rho(t)) ) =
\cos^2 (\psi - \psi')$, and it suffices to write the response function
directly in terms of $\psi$ and $\psi'$, although we emphasize that
more complex detector angular response functions are possible.
Substituting these expressions into equation (2) gives,
\begin{equation}
O(E, \psi) = \int_{E'} \int_{\psi'} \left (h(E') + g(E')\delta(\psi' -
\psi'_0(E')) \right ) \epsilon(E') R(E',E) C_0 ( A(E') + B(E')\cos^2 (\psi -
\psi'(E')))dE' d\psi' \; .
\end{equation}
For the term proportional to the unpolarized component, $h(E')$, we can 
rewrite the integral as,
\begin{equation}
\int_{E'} \frac{h(E')\epsilon(E')R(E',E)}{\pi(A(E') + B(E')/2)} \int_{\psi'}
\left [ A(E') + B(E')\cos^2 (\psi - \psi'(E')) \right ] d\psi' dE' \;,
\end{equation}
where we have explicitly included $C_0$ given above. The integration
over $\psi'$ is now straightforward, and is simply equal to $\pi(A(E')
+ B(E')/2)$, thus for this term the $\psi$ dependence integrates out.
This is as expected, since for the assumed form of $V(E',\psi',\psi) =
v(E', (\psi - \psi'))$, where $\mu(E')$ is a function of $E'$ only
(and not $\psi'$), an unpolarized source produces a uniform
distribution in observed angle $\psi$.  We emphasize that if this is
not the case, and the modulation fraction is also a function of the
intrinsic polarization angle, $\psi'$, then one should not expect an
unpolarized source to produce a uniform (flat) angular distribution in
observed angle $\psi$ (we discuss this further in \S 4).  This term
can then be written as,
\begin{equation}
\int_{E'} h(E')\epsilon(E')R(E',E) dE' \; .
\end{equation}
Thus, as intuition would suggest, the unpolarized term looks exactly
like the analogous case with no polarization sensitivity.

The remaining term involving the polarized spectrum, $g(E')$, is,
\begin{equation}
\int_{E'} \int_{\psi'} \frac{g(E')\delta(\psi' -
  \psi'_0(E'))\epsilon(E')R(E',E)}{\pi(A(E') + B(E')/2)} \left [ A(E') +
  B(E')\cos^2 (\psi - \psi'(E')) \right ] d\psi' dE' \;.
\end{equation}
In this case the angular integration is simplified by the delta
function, which restricts the source polarized photons to have the
intrinsic angle $\psi'_0(E')$. This reduces the integral to,
\begin{equation}
\int_{E'} \frac{g(E')\epsilon(E')R(E',E)}{(A(E') + B(E')/2)} \left [ A(E') +
  B(E')\cos^2 (\psi - \psi'_0(E')) \right ]  dE' \;,
\end{equation}
where we have picked up a factor of $\pi$ from the delta function
integration.  Now, using the definition of $\mu(E')$ above it can be
shown that
\begin{equation}
\frac{A(E')}{(A(E') + B(E')/2)} = 1 - \mu(E') \; ,
\end{equation}
and 
\begin{equation}
\frac{B(E')}{(A(E') + B(E')/2)} = 2 \mu(E') \; .
\end{equation}
With these substitutions we can re-write equation (12) in the form,
\begin{equation}
\int_{E'} g(E')\epsilon(E')R(E',E) \left [ (1 - \mu(E')) +
  2\mu(E')\cos^2(\psi - \psi'_0(E')) \right ] dE' \; .
\end{equation}
Combining all the terms from equation (8) we have the result,
\begin{equation}
O(E,\psi) = \int_{E'} ( ( h(E') + g(E')(1-\mu(E'))) \epsilon(E')R(E',E) \;
\end{equation}
\begin{equation}
+ 2g(E')\mu(E')\epsilon(E')R(E',E)\cos^2(\psi - \psi'_0(E')) )  ) dE' \; . 
\end{equation}
This integrand is of the form $\alpha(E',E) + \beta(E',E)\cos^2(\psi -
\psi'_0(E'))$ if we make the identifications, $\alpha(E',E) = (h(E') +
g(E')(1-\mu(E')))\epsilon(E')R(E',E)$, and $\beta(E',E) =
2g(E')\mu(E')\epsilon(E')R(E',E)$. Now, with the help of some
trigonometric identities it is straightforward to show that,
\begin{equation}
\alpha(E',E) + \beta(E',E)\cos^2(\psi - \psi'_0(E')) = \left (
(\alpha(E',E) + \frac{\beta(E',E)}{2} \right ) 
\end{equation}
\begin{equation}
+ \left ( \frac{\beta(E',E)}{2}\sin(2\psi'_0(E')) \right ) \sin(2\psi)
+ \left ( \frac{\beta(E',E)}{2}\cos(2\psi'_0(E')) \right ) \cos(2\psi) 
\end{equation}
(see, for example, Strohmayer \& Kallman 2014).  If we further define
$I(E',E) = (\alpha(E',E) + \beta(E',E)/2)$, $U(E',E) =
(\beta(E',E)/2)\sin(2\psi'_0(E'))$, and $Q(E',E) =
(\beta(E',E)/2)\cos(2\psi'_0(E'))$, then equation (16) can be expressed
in the familiar Stokes form
\begin{equation}
O(E, \psi) = I(E',E) + U(E',E)\sin(2\psi) + Q(E',E)\cos(2\psi) \; ,
\end{equation}
with, 
\begin{equation}
I(E) = \int_{E'} ( h(E') + g(E')(1-\mu(E')) +
g(E')\mu(E') ) \epsilon(E')R(E',E)dE'
\end{equation}
\begin{equation}
= \int_{E'} (h(E') + g(E'))\epsilon(E')R(E',E)dE' 
\end{equation}
\begin{equation}
U(E) = \int_{E'} g(E') \mu(E')\epsilon(E')R(E',E)\sin(2\psi'_0(E')) dE' \; ,
\end{equation}
and 
\begin{equation}
Q(E) = \int_{E'} g(E') \mu(E')\epsilon(E')R(E',E)\cos(2\psi'_0(E')) dE'\; , 
\end{equation}
where now we have explicitly included the integration over $E'$.
These equations relate the intrinsic source properties (spectral and
linear polarization properties defined by $h(E')$, $g(E')$ and
$\psi'_0(E')$) to the modulation curve observed by a polarization
sensitive detector characterized by three response functions, the
traditional energy response functions $\epsilon(E')$ (effective area)
and $R(E',E)$ (energy redistribution matrix), and the energy dependent
modulation fraction $\mu(E')$, which encompasses the detector's
polarization sensitivity. With the help of equations (6) and (7) these
can also be written as,
\begin{equation}
I(E) = \int_{E'} F(E')\epsilon(E')R(E',E)dE' 
\end{equation}
\begin{equation}
U(E) = \int_{E'} W(E') \mu(E')\epsilon(E')R(E',E) dE'
\end{equation}
and 
\begin{equation}
Q(E) = \int_{E'} Z(E') \mu(E')\epsilon(E')R(E',E) dE' \; .
\end{equation}

In thinking further about modeling X-ray sources including their
linear polarization properties, the above discussion outlines a
path. An observation in a particular energy channel, $E$ (or range of
channels), is a background-subtracted counts or count rate modulation
curve of the form,
\begin{equation} 
O(E, \psi) = I(E) + U(E)\sin(2\psi) + Q(E)\cos(2\psi) \; .
\end{equation} 
One can perform a $\chi^2$ fit to the observed modulation curve for
each energy channel, producing the three observed Stoke's spectra,
with their associated uncertainties.  One can then define source
models, using, for example, parameterizations for $F(E')$, $a(E')$,
and $\psi'_0(E')$ (recall that the polarized spectrum $g(E') =
F(E')a(E')$), generate predicted spectra using equations (25), (26)
and (27), and then carry out $\chi^2$ minimization to find the source
parameters which best fit the observed spectra in a statistical sense.
This is entirely analogous to the simpler case with no polarization
sensitivity, except that the generalization to spectro-polarimetry
requires the joint fitting of three observed spectra, one for each of
the Stokes parameters.  One model spectrum, $F(E')$, is folded through
the full detector response function, $\epsilon(E')R(E', E)$, and the
two new spectra, $W(E')=F(E')a(E')\sin(2\psi'_0(E'))$ and
$Z(E')=F(E')a(E')\cos(2\psi'_0(E'))$ are folded through the
``modulated response'' function, $\mu(E')\epsilon(E')R(E',E)$.

The forward fitting procedure implemented by XSPEC can in principle
accommodate this process with a few simple additions. For example,
$F(E')$ is a physical energy spectrum, and the XSPEC package includes
many such options. In order to compute the model-predicted spectra
$U(E)$ and $Q(E)$ one would require model parameterizations for
$a(E')$ and $\psi'_0(E')$. Such models do not yet exist in the current
XSPEC implementation, but they could be easily added. In XSPEC
parlance they would be relatively simple multiplicative model
components. Further, the additional detector modulation function,
$\mu(E')$, is very much like an effective area function, which can be
included in the XSPEC implementation as a so-called ``ancillary
response function'' (an ``arf'' file), so this could easily be
incorporated in the same way.

\section{Detector Considerations}

In the discussion above we assumed that the detector's angular
response function satisfied the condition that $V(E',\phi', \phi) =
v(E',(\phi - \phi'))$, that is, we assumed that for a given photon
energy $E'$ all intrinsic polarization angles, $\phi'$, produce the
same modulation fraction. This need not necessarily be the case. A
simpler way to say this is that a detector system could in principle
measure some intrinsic polarization angles better than others.

Photoelectric polarimeters work by imaging the charge track of a
photoelectron produced when an X-ray photon is absorbed in the
detection gas (see Costa et al. 2001; Black et al. 2004). The charge
track must be drifted some distance and then detected in a pixellated
detector/readout system.  Detectors of this type have been developed
in two basic geometries.  One, which we call an ``imaging''
polarimeter, drifts the charge track in the same direction as the
incident photon beam (Costa et al. 2001).  The other, known as a
``time projection chamber'' (TPC) polarimeter, drifts the charge in a
direction orthogonal to the photon beam (Black et al. 2004).  When a
charge track drifts it also diffuses, smearing out the track.  If it
drifts too far before being detected diffusion will completely erase
the directional (polarization) information within the track.  In order
to keep diffusion to a reasonable level an imaging polarimeter must
have a relatively shallow layer of detection gas (unless diffusion can
be reduced in some other fashion). The pixellated readout system makes
up the bottom of this layer.  This provides a limit to the efficiency
of detectors constructed in this geometry.  By contrast, a TPC
polarimeter can contain a much greater depth of absorbing gas (and
therefore have a greater efficiency, other things being equal), since
the tracks are drifted to the side of the detection volume.  However,
the position of the track within the field of view is better-sampled
in the imaging geometry, so by centroiding the track or reconstructing
it, one can estimate where it ocurred on the sky (hence the imaging
appellation). Since the track drifts to the side in a TPC the location
information regarding where it interacted within the volume is at
least partially lost.

Several effects in the photoelectron track imaging and reconstruction
process can, in principle, result in some intrinsic angles producing
higher modulations.  For example, the detector readout formats have
some intrinsic, pixellated geometry. Some employ either hexagonal or
Cartesian geometries.  Since the track image is generally of modest
resolution, angles corresponding to symmetries of the underlying
readout geometry could in principle be better resolved than
others. This could result in particular intrinsic angles producing a
higher modulation than others.

Another possible cause of variation in the modulation with intrinsic
angle results from potential drift asymmetries.  Since tracks that
drift for longer have a greater time to diffuse, the longer a track
drifts the poorer, on average, is the accuracy with which the track
angle can be measured.  For one thing this means that the modulation
fraction in such a detector is a function of the drift distance. Since
X-rays are absorbed over a range of drift distances, the detector's
angular response will be an average over the modulation as a function
of drift distance, weighted by the relative number of photons absorbed
at each distance. Since the average depth at which an X-ray is
absorbed is also a function of X-ray energy, the total distribution of
drift distances will depend to some level on the intrinsic photon
energy.  As noted above, these effects are included in any
experimentally determined angular response function.

Considering a single interaction point for simplicity, there is a
potential drift asymmetry introduced by the geometry of a TPC compared
to an imaging polarimeter.  In an imaging polarimeter the
photoelectrons are preferentially ejected in a plane that is parallel
to the detector plane.  Thus, for the imaging polarimeter, most of the
tracks are drifted for the same distance.  This symmetry is broken in
the TPC geometry, since the photoelectrons are now preferentially
ejected in a plane which is perpendicular to the detection plane.  In
this case half the tracks are ejected toward the detection plane and
half away from it. Those directed at the detection plane will travel
for a somewhat shorter distance (time), and will suffer slightly less
diffusion than those directed away from the detection plane.  If we
measure track angles from 0 to $2\pi$ with $0$ (and $2\pi$)
representing tracks directed straight at the detection plane and $\pi$
those directed straight away from the detection plane, then those with
intrinsic angle of zero diffuse slightly less on average than those
with angle $\pi$.  Thus, the charge tracks with angle $0$ can be
expected to be slightly ``sharper'' (better resolved) and therefore
their angles somewhat more accurately determined.  While it is beyond
the scope of this paper to explore such effects in detail, the
discussions above establish that it should not be totally unexpected
for a photoelectric polarimeter to be described by an angular response
function that does not exactly satisfy $V(E',\phi', \phi) = v(E',
(\phi - \phi'))$, although in many circumstances the assumption of
uniformity is likely to be a very good one.

Nevertheless, it is instructive to explore the non-uniorm response
case a bit further, and below we will show that in such a case the
response to an unpolarized source does not produce a flat modulation
curve.  We can demonstrate this by introducing a slightly more general
response function. First, the properly normalized ``uniform'' response
function used above can be expressed as,
\begin{equation}
V(E', \phi', \phi) = (1/\pi) \left ( (1 - \mu(E')) + 2\mu(E')\cos^2
(\phi - \phi') \right ) \; .
\end{equation}
We can introduce a simple ``non-uniform'' response by defining
$\mu(E', \phi') = \mu(E')\eta(\phi')$, where $\eta$ is a slowly
varying function of $\phi'$. We can mimic an illustrative non-uniform
effect as described above by defining $\eta = 1 - (2d/\pi)|\phi'|$.
Here, $d$ is just a small constant factor that defines the size of a
linear change in the modulation as $\phi'$ ranges from $0$ to $\pm
\pi/2$. In this case the modulation would be a maximum for $\phi' = 0$
and drop to $1 - d$ at $\phi' = \pm \pi/2$. Figure 1 compares the
response functions in these uniform and non-uniform cases. The black
curves show uniform response functions (with $\mu(E') = 0.5$) for 5
equally spaced values of $\phi'$.  One can see that the peaks shift
for different $\phi'$ values, but the maximum values of the response
are equal (as expected for a uniform response). The dashed black line
is equal to the integral of the uniform response over $\phi'$ (minus
0.68 to plot it within the same $y$ range, the integral is unity by
definition), and is flat as expected. The red curves in Figure 1 show
a non-uniform response function for the same $\phi'$ values, and with
$\mu=0.5$ and $d = 0.3$. It is fairly easy to see how the modulation
amplitude drops as $\phi'$ moves away from 0. The dashed red curve
shows the integral over all $\phi'$ for the non-uniform response
function (again minus 0.68 for plotting purposes), and it is evidently
not flat, but peaks where the response shows the largest modulation
amplitude.  It should be emphasized that the value of $d$ chosen in
this case is purely for illustrative purposes only, and is not meant
to represent any particular detector system.  

While some might consider such an effect a detector ``systematic,''
this is really not an accurate description, as it is simply a part of
``how the detector works,'' and one can still generate predicted
modulation curves, and do spectro-polarimetry based on equation (2),
though, depending on the complexity of an actual non-uniform response,
the resulting observed modulation curves may not have a simple
analytical representation, and numerical evaluation of the angular
integrals in (2) could be necessary.  In any case, the problem is
still well defined mathematically.  It should be noted that in the
above discussion we were considering the observed response in a
detector's frame.  If the detector is fixed (not rotating), then a
similar response to unpolarized flux would be evident in the sky frame
as well, though perhaps with some constant offset in angle specified
by the mapping from detector to sky frames.

Since polarization properties have not been measured for most
astrophysical sources, a commonly raised concern is that a
``non-uniform'' response to unpolarized flux might be falsely claimed
as a significant polarization measurement.  But this again assumes
that all polarimeters will produce flat modulation curves when
illuminated with unpolarized flux, and this is only true if the
response is uniform as described above. The key issue, as with any
observational claims, is that they be based on accurate and reliable
instrument calibrations.  Thus, polarimeters with non-uniform
responses can also be effective instruments, however, the potential
risk that a non-uniformity could lead to a false polarization claim
argues for more careful attention to detector modeling, calibration
and monitoring in such a case.

While we emphasize that the best way to mitigate against a false
polarization claim is a proper understanding of ones detector response
function, one can also ``enforce'' flatness of the response to
unpolarized flux in the sky frame by rotation of the detector about
the line of sight. As outlined earlier, rotation of the detector
provides a mapping from sky angle to detector frame angle of the form
$\psi' = \phi' + \Omega t$, where $\Omega$ is the angular rotation
rate of the spacecraft.  Spacecraft rotation tends to enforce the
condition that all intrinsic sky angles are measured at all intrinsic
detector frame angles. This has the effect of ``smoothing out'' a
non-uniform response to unpolarized flux of the kind described in \S 4
(see Figure 1), when plotted in sky coordinates, and as long as the
source specific flux and polarization variability time scales are much
longer than the spacecraft rotation period, and sufficient exposure
time is achieved at all rotation angles.  Thus, in this case one can
always define an effectively uniform response function by carrying out
the angular average created by rotation of the detector frame.

\section{Simulated Observations and Data Modeling}

We now walk through an example of spectro-polarimetric fitting using
simulated observations with parameters and response functions
appropriate for observations with the proposed {\it PRAXyS} Small
Explorer mission (Iwakiri et al. 2016). As is appropriate in this
case, we consider a uniform angular response function, such that the
modulation function, $\mu$, is only a function of intrinsic photon
energy, $E'$ (see eqn 29).

We first define a source model using the total spectrum, $F(E')$, the
polarization amplitude, $a(E')$, and the polarization position angle,
$\psi_0'(E')$.  As an illustrative example we choose model parameters
consistent with the known spectrum and polarization properties of the
Crab nebula (see Weisskopf et al. 1978). We use a power-law photon
spectrum, $F(E') = C_n E'^{-\alpha}$, with index $\alpha = 2.1$. For
the polarization properties we assume simple linear dependencies with
energy. We take $a(E') = a_0 + \Delta a (E' - 2.6 \; {\rm keV})$ and
$\psi'_0(E') = \psi'_0 - \Delta\psi'_0 (E' - 2.6 \; {\rm keV})$. We
use $a_0 = 0.19$ and $\psi'_0 = 156$ deg, which are approximately
consistent with the measured values at $2.6$ keV, but we allow for
small linear changes with $E'$ to explore the sensitivity of a {\it
  PRAXyS} observation to such changes. For this purpose we take
$\Delta a = 0.01$ keV$^{-1}$ and $\Delta\psi'_0 = 1$ deg keV$^{-1}$. 

For the response functions we use an effective area curve and quantum
efficiency appropriate for the proposed {\it PRAXyS} Small Explorer
(SMEX) observatory, and we use a modulation function, $\mu(E')$ based
on measurements obtained with and detector simulations of the {\it
  PRAXyS} polarimeter (Iwakiri et al. 2016). For this demonstration
example we employ a simplified diagonal redistribution matrix, and
since the Crab is a bright source we ignore backgrounds.  Figure 2
shows the resulting full (black) and modulated (red) effective areas
(in units of cm$^2$) used in our simulations, as well as the energy
dependent modulation function, expressed as a percentage (green).

To simulate an observation we carry out the following procedures. We
assume a constant source count rate equal to that expected from the
Crab nebula, and first assign a random event time.  We take the total
spectrum and fold it through the full response matrix, obtaining a
predicted count rate spectrum in energy channel space.  We then
convert that to a cumulative distribution function and use the
so-called transformation method to make random energy channel
draws. Once we have the energy of the photon, we draw a random deviate
between $0$ and $1$ and use $a(E')$ defined above to assign it to the
``polarized'' or ``unpolarized'' angular distributions the correct
fraction of times. For example, if it is ``polarized,'' then we assign
it the correct sky position angle, $\psi'_{0}(E')$, for it's energy.
If it's ``unpolarized,'' we assign it a sky position angle that is a
random draw from a uniform distribution of sky angles.  Finally, we
``detect'' the event's observed sky angle using the appropriate
angular response function for its energy (see equation 29). In this
way we can build up a simulated observation of $N$ events.

We also allow for the possibility of uniform angular rotation of the
spacecraft.  With rotation included a few additional steps are
required for the simulations. First, after assigning an event's
intrinsic sky angle we find the corresponding detector angle for this
sky angle based on the event time and the mapping from sky to detector
angle. Then, we assign an observed detector angle using the
appropriate angular response (modulation) function, and then, finally,
we place that detector frame angle back on the sky (detected sky
angle) using the same mapping.

\subsection{Results of Crab Nebula Simulation}

The proposed baseline science plan for {\it PRAXyS} would result in
millions of counts detected from the Crab nebula and pulsar, so we
illustrate the capabilities and methods with a simulation for 6
million detected counts from the nebula.  For the $\chi^2$ fitting we
first group the energy bins such that each new channel has at least
$1.5 \times 10^5$ counts.  For each energy channel we then determine
the three Stokes parameters that describe its observed modulation
curve.  This can be done in two effectively equivalent ways.  One is
to define $M$ angular bins and then bin up the events into a
modulation curve and do $\chi^2$ fitting (see, for example, Strohmayer
\& Kallman 2014).  However, in this case one can also use a bin-free
estimator for the Stokes parameters, such that, $I(E) = N_E$, $U(E) =
2 \sum_{i=1}^{N_E} \sin(2\psi_i)$, and $Q(E) = 2 \sum_{i=1}^{N_E}
\cos(2\psi_i)$ (see, for example, Montgomery \& Swank 2015; Kislat et
al. 2015b). Here, $N_E$ is just the total number of events in energy
channel $E$.  For $I(E)$ the uncertainty is simply $\sigma_I =
(N_E)^{1/2}$, and for $Q(E)$ and $U(E)$ we use $\sigma_Q = \sigma_U =
(2N_E)^{1/2}$.  We used both estimates and found they give consistent
results. Here we present results using the bin-free method.

With the three observed Stokes energy spectra we can now carry out
$\chi^2$ fitting to constrain the parameters of the
spectro-polarimetric model described above. We use equations (22),
(23) and (24) to determine the model-predicted spectra, and we jointly
fit the simulated data to all three spectra. We use a spectral model
written in IDL along with a least-squares fitting routine also
developed within IDL that is based on MINPACK-1 (Markwardt 2009). The
model has six free parameters, two each for the power-law energy
spectrum, $F(E')$, the polarization amplitude, $a(E')$, and the
polarization position angle, $\psi'_0(E')$. Figures 3 through 8
summarize results of the simulation with the parameters described
above.  Figures 3 and 4 show the resulting Stokes spectra, I(E)
(Figure 3), $Q(E)$ and $U(E)$ (Figure 4), along with the best fitting
model spectra (solid curves running through the data points).  In
addition to this we also plot the difference between the data and best
fitting models. For this example we have 84 total spectral bins (28
for each Stokes spectrum) and 6 free parameters for a total of 78
degrees of freedom.  We find an acceptable minimum $\chi^2$ value of
77.6.

Figure 5 shows a ``residuals'' plot of $\chi = (\rm{Data} - {\rm
  Model}) / \sigma_{\rm data}$.  The inferred fractional polarization
amplitude (top) and position angle (bottom) versus energy are shown in
Figure 6. Here, the top panel shows the observed modulation amplitude
in each energy channel, $a(E) = (Q(E)^2 + U(E)^2 )^{1/2}/I(E)$ (red
points), and the data points (black square symbols) show the inferred
polarization amplitude, $a_p(E) = a(E)/ \mu_{avg}(E)$ with $1\sigma$
error bars. Here, $\mu_{avg}(E)$ is the mean modulation in each
grouped energy channel. The inferred polarization position angle
versus energy, $\psi'(E) = 1/2 \tan^{-1} (Q(E)/U(E))$, is shown in the
bottom panel.  The solid (black) line in each panel is the best
fitting polarization model (either amplitude or position angle), and
the dashed lines show the input models used to generate the simulation
(the ``true'' model).  The red symbols in both panels show the
existing polarization measurements of the Crab nebula as an indication
of the current state of knowledge (Weisskopf et al. 1978).  Finally,
in Figures 7 and 8 we show the derived confidence regions for both the
polarization amplitude parameters (Figure 7) and the position angle
(Figure 8). We show contours drawn at $\Delta\chi^2 = 2.3$ and $4.61$,
which correspond to confidence levels of $68.3$ and $90$ \%. The green
square symbols mark the input (``true'') values. These results
demonstrate that the spectro-polarimetric forward fitting procedure
recovers the input model parameters within the expected statistical
precision.  Existing polarization measurements of the Crab nebula are
consistent with no variation in either fractional amplitude or
position angle with photon energy. The results of the simulations
described above indicate that a mission like {\it PRAXyS} would be
extremely sensitive to such variations.  For example, Figures 7 and 8
indicate that with 6 million detected photons, variations in
fractional amplitude at the level of 1\% keV$^{-1}$, and position
angles at the level of 1 deg keV$^{-1}$, can clearly be detected.

%

\subsection{Summary and Conclusions}

We have presented a generalization of the standard ``forward fitting''
procedure for X-ray spectroscopy to include linear polarization of
X-ray sources. When the angular response of the polarimeter is
``uniform,'' in the sense that for a given photon energy all intrinsic
photon polarization angles produce the same fractional modulation,
then the polarization sensitivity introduces two additional observed
spectra, related to the Stokes $U(E)$ and $Q(E)$ parameters. Thus,
joint fitting of three observed spectra can yield constraints on
spectro-polarimetric source models. The computation of the predicted
spectra as a convolution of the source spectral model with the
detector energy response function maintains the same familiar form,
however, for the new $U(E)$ and $Q(E)$ energy spectra the appropriate
detector response function is the ``modulated'' response,
$\mu(E')\epsilon(E') R(E',E)$, which is just the traditional energy
response function multiplied by the detector's energy-dependent
modulation function, $\mu(E')$. The additional functionality required
for spectro-polarimetry is thus relatively straightforward, and could
be incorporated within exiting X-ray spectral software tools, as for
exaxmple, XSPEC, with relatively simple modifications.

Several previous studies have also explored aspects of X-ray spectral
analysis in the context of polarimetry.  For example, Kislat et
al. (2015a) described an iterative ``unfolding'' method based on
Bayes' Theorem to obtain model-independent estimates of the energy
spectrum and polarization properties, and they presented simulated
results with this method appropriate for the {\it X-Calibur} hard
X-ray Compton scattering polarimeter. In addition, Krawczynski (2011)
explored a maximum likelihood analysis method for Compton polarimeters
based on measuring both the azimuthal and polar angles of the
scattered photons. There are similarities between these methods and
the forward-folding procedure we describe here. For example, they both
account for energy-dependent effects with a multidimensional response
function that models how input and output observables (such as energy
and position angle) are related through the detection process. The
iterative procedure appears to be more computationally intensive, but
in principle, returns model-independent estimates of source spectra
and polarization properties.  On the other hand, forward folding can
likely be more easily incorporated into existing software tools (such
as XSPEC), and while not strictly model-independent, it still enables
important insights regarding source properties.

While it is beyond the scope of this paper to derive results
appropriate for all currently proposed X-ray polarimeters, the basic
methods discussed here should also be applicable to instruments
working in the hard X-rays and gamma-rays.  However, it is possible
that additional observables may need to be included in the response
functions. For example, in addition to measuring the energy and
azimuthal scattering angles, the hard X-ray scattering experiments,
such as {\it X-Calibur}, benefit from also measuring the polar
scattering angle (Krawczynski 2011). Moreover, issues associated with
uniformity of the response functions, as discussed in \S 4, would have
to be explored for specific detector systems.

\acknowledgements

The author acknowledges helpful discussions with Keith Jahoda, Craig
Markwardt, Tim Kallman, and Jean Swank.  We also thank the anonymous
referee for a helpful review of this paper.

\newpage

\section*{References}

\begin{enumerate}

\item[]{} Sanchez Almeida, J., \& Martinez Pillet, V.\ 1993, \ao, 32,
  4231

\item[]{} Arnaud, K.~A.\ 1996, Astronomical Data Analysis Software and
  Systems V, 101, 17

\item[]{} Bannister, N.~P., Harris, K.~D.~M., Collins, S.~P., et
  al.\ 2006, Experimental Astronomy, 21, 1

\item[]{} Beilicke, M., Cowsik, R., Dowkontt, P., et al.\ 2014,
  Multifrequency Behaviour of High Energy Cosmic Sources, 293

\item[]{} Bellazzini, R., Angelini, F., Baldini, L., et al.\ 2003,
  \procspie, 4843, 372

\item[]{} Black, J.~K., Deines-Jones, P., Hill, J.~E., et al.\ 2010,
  \procspie, 7732, 25

\item[]{} Black, J.~K., Deines-Jones, P., Jahoda, K., Ready, S.~E., \&
  Street, R.~A.\ 2004, \procspie, 5165, 346

\item[]{} Chauvin, M., Jackson, M., Kawano, T., et al.\ 2016,
  Astroparticle Physics, 82, 99

\item[]{} Chipman, R.~A., Brown, D.~M., \& McGuire, J.~P., Jr.\ 1993,
  \ao, 32, 4236

\item[]{} Chipman, R.~A., Brown, D.~M., \& McGuire, J.~P., Jr.\ 1992,
  \ao, 31, 2301

\item[]{} Collins, S.~P.\ 1997, Nuclear Instruments and Methods in
  Physics Research B, 129, 289

\item[]{} Costa, E., Soffitta, P., Bellazzini, R., et al.\ 2001, \nat,
  411, 662

\item[]{} Hamaker, J.~P., \& Bregman, J.~D.\ 1996, \aaps, 117, 161

\item[]{} Hill, J.~E., Black, J.~K., Jahoda, K., et al.\ 2016,
  \procspie, 9905, 99051B

\item[]{} Hill, J.~E., Black, J.~K., Emmett, T.~J., et al.\ 2014,
  \procspie, 9144, 91441N

\item[]{} IAU, 1974, Transactions of the IAU Vol. 15B (1973) 166

\item[]{} Iwakiri, W.~B., Black, J.~K., Cole, R., et al.\ 2016,
  Nuclear Instruments and Methods in Physics Research A, 838, 89

\item[]{} Jahoda, K., Kallman, T.~R., Kouveliotou, C., et al.\ 2016,
  \procspie, 9905, 990516

\item[]{} Kislat, F., Beilicke, M., Guo, Q., Zajczyk, A., \&
  Krawczynski, H.\ 2015a, Astroparticle Physics, 64, 40

\item[]{} Kislat, F., Clark, B., Beilicke, M., \& Krawczynski,
  H.\ 2015b, Astroparticle Physics, 68, 45

\item[]{} Krawczynski, H.~S., Stern, D., Harrison, F.~A., et
  al.\ 2016, Astroparticle Physics, 75, 8

\item[]{} Krawczynski, H.\ 2011, Astroparticle Physics, 34, 784

\item[]{} Lampton, M., Margon, B., \& Bowyer, S.\ 1976, \apj, 208, 177

\item[]{} Markwardt, C.~B.\ 2009, Astronomical Data Analysis Software
  and Systems XVIII, 411, 251

\item[]{} Montgomery, C.~G., \& Swank, J.~H.\ 2015, \apj, 801, 21

\item[]{} Soffitta, P., Bellazzini, R., Bozzo, E., et al.\ 2016,
  \procspie, 9905, 990515

\item[]{} Strohmayer, T.~E., \& Kallman, T.~R.\ 2013, \apj, 773, 103

\item[]{} Weisskopf, M.~C., Ramsey, B., O'Dell, S., et al.\ 2016,
  \procspie, 9905, 990517

\item[]{} Weisskopf, M.~C., Silver, E.~H., Kestenbaum, H.~L., Long,
  K.~S., \& Novick, R.\ 1978, \apjl, 220, L117

\end{enumerate}

\newpage

\begin{figure}
\epsscale{0.85}
\plotone{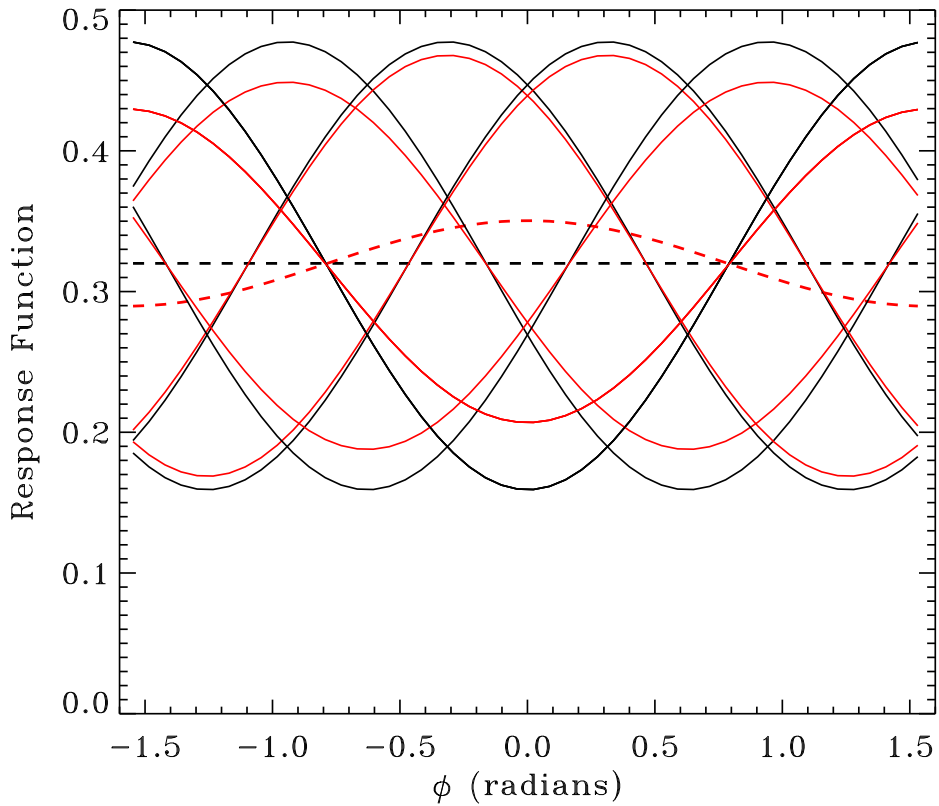}
\caption{Angular response functions for responses which are
  independent of the intrinsic detector frame angle (black, see
  equation (29)), and for which the modulation fraction has a simple
  linear dependence on intrinsic angle $\phi'$ (red). In each case
  curves are shown for 5 equally spaced values of $\phi'$. The red
  curves show a non-uniform response function for the same $\phi'$
  values, and with $\mu=0.5$ and $d = 0.3$. The dashed curves show the
  integral over all $\phi'$ for both the uniform (black) and
  non-uniform (red) response functions (minus 0.68 to fit on the same
  vertical scale). The non-uniform response is evidently not flat, but
  peaks where the response shows the largest modulation amplitude.
  See the discussion in \S 4 for more details.  \label{fig1}}
\end{figure}

\newpage

\begin{figure}
\epsscale{0.85}
\plotone{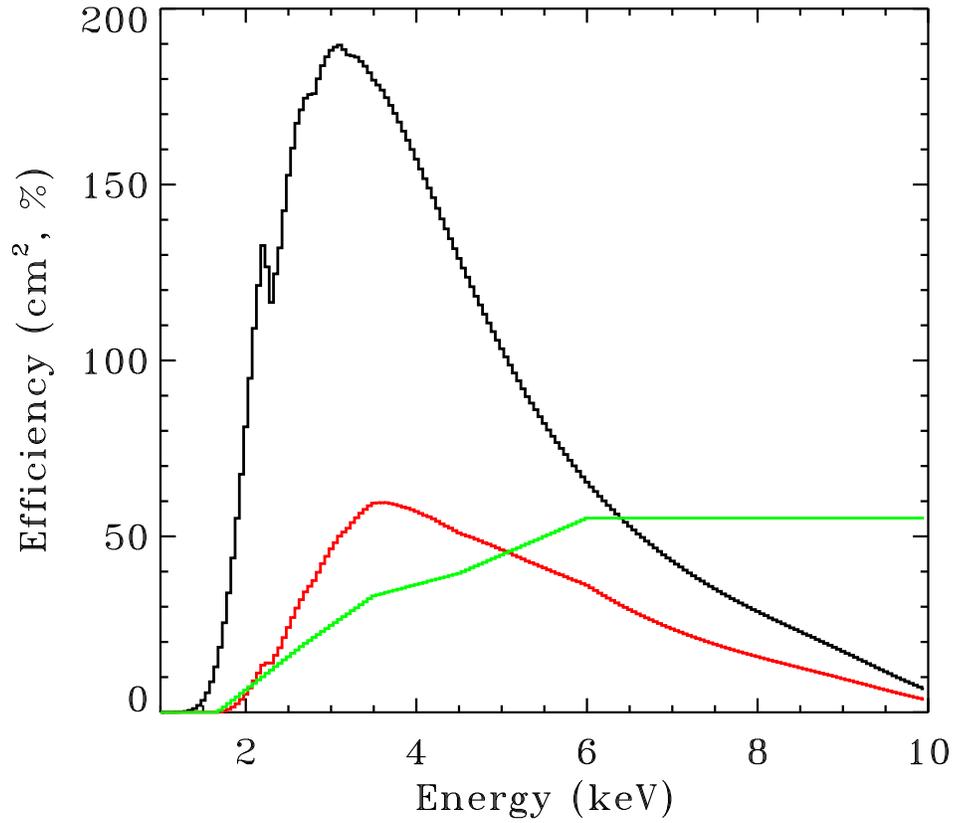}
\caption{Effective area curves for the full energy response function
  (black), and the ``modulated'' response (red), used in the
  spectro-polarimetric fitting simulations. We also show the assumed
  energy dependent modulation function, $\mu(E')$ (green) expressed as
  a percentage.  \label{fig1}}
\end{figure}

\newpage

\begin{figure}
\epsscale{0.85}
\plotone{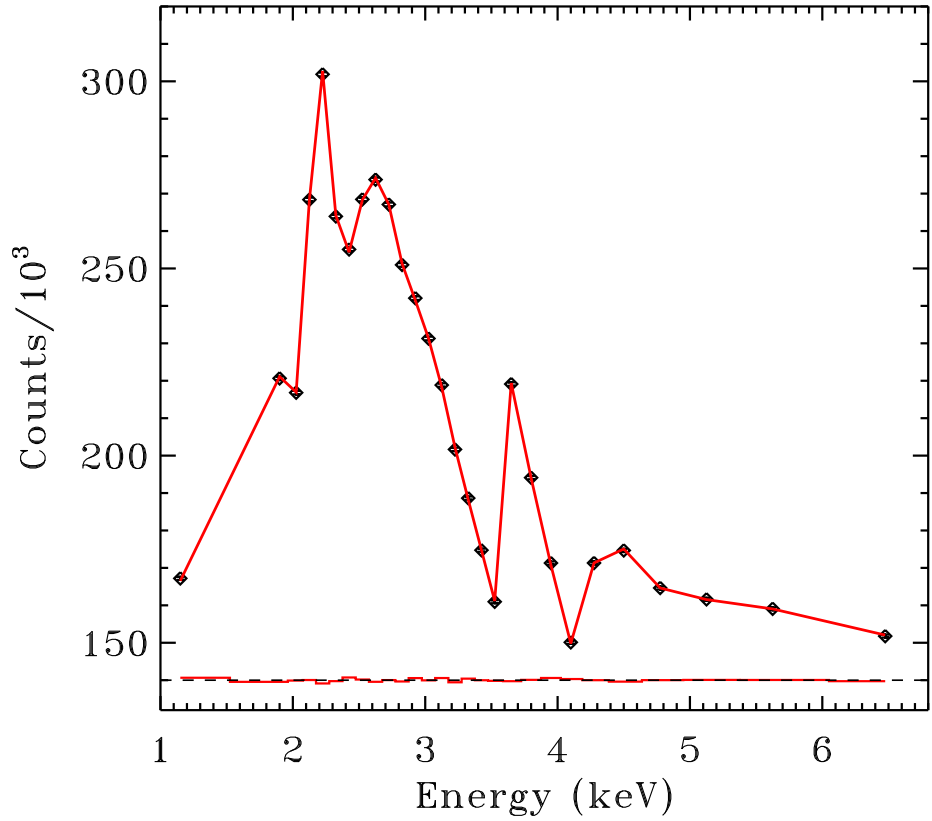}
\caption{Stokes spectrum $I(E)$ for our 6 million count Crab nebula
  simulation described in \S 5.  The black symbols and error bars show
  results of the simulated data, and the solid red curve is the best
  fitting model for Stokes $I(E)$. The horizontal red curve running
  through zero is the data minus the best-fitting model. See the
  discussion in \S 5 for additional details.  \label{fig1}}
\end{figure}

\newpage

\begin{figure}
\epsscale{0.85}
\plotone{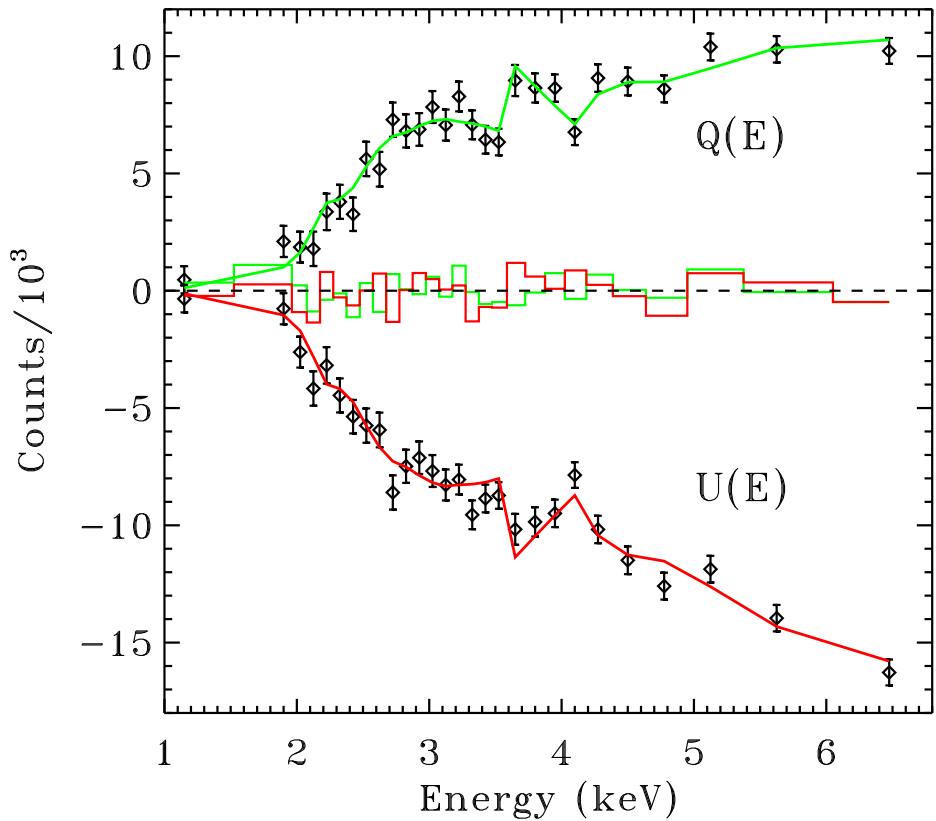}
\caption{Stokes spectra $U(E)$ (lower) and $Q(E)$ (upper) for our 6
  million count Crab nebula simulation described in \S 5.  The black
  symbols with error bars show the simulated data, and the solid
  curves show the best-fitting models for $U(E)$ (red) and $Q(E)$
  (green). The green and red histograms running through zero are the
  data minus the model for each spectrum. See the discussion in \S 5
  for additional details. \label{fig1}}
\end{figure}
\newpage

\begin{figure}
\epsscale{0.85}
\plotone{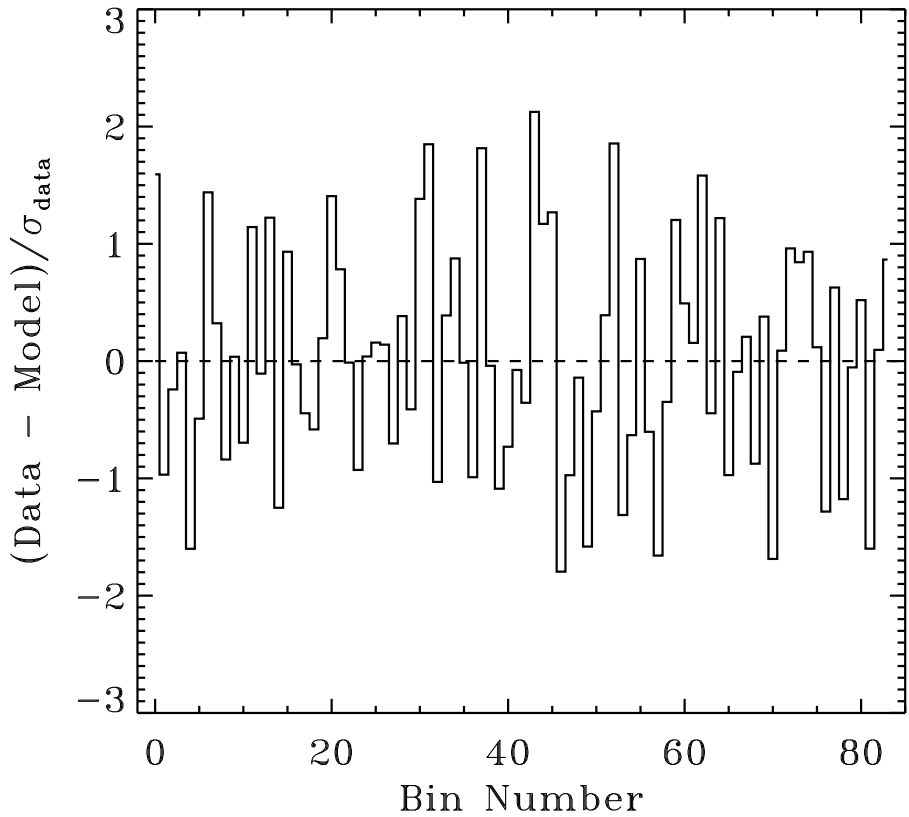}
\caption{Residuals defined as $\chi = ({\rm Data} - {\rm
    Model})/\sigma_{\rm data}$ for the best-fitting model for our 6
  million count Crab nebula simulation described in \S 5. The fit to
  all three spectra are shown in terms of ``bin number,'' with 28
  bins for each spectrum in the order $I(E)$, $U(E)$ and $Q(E)$. For
  example, bins 0 - 27 correspond to the residuals for Stokes
  $I(E)$. See the discussion in \S 5 for additional
  details.  \label{fig1}}
\end{figure}

\newpage

\begin{figure}
\epsscale{0.85}
\plotone{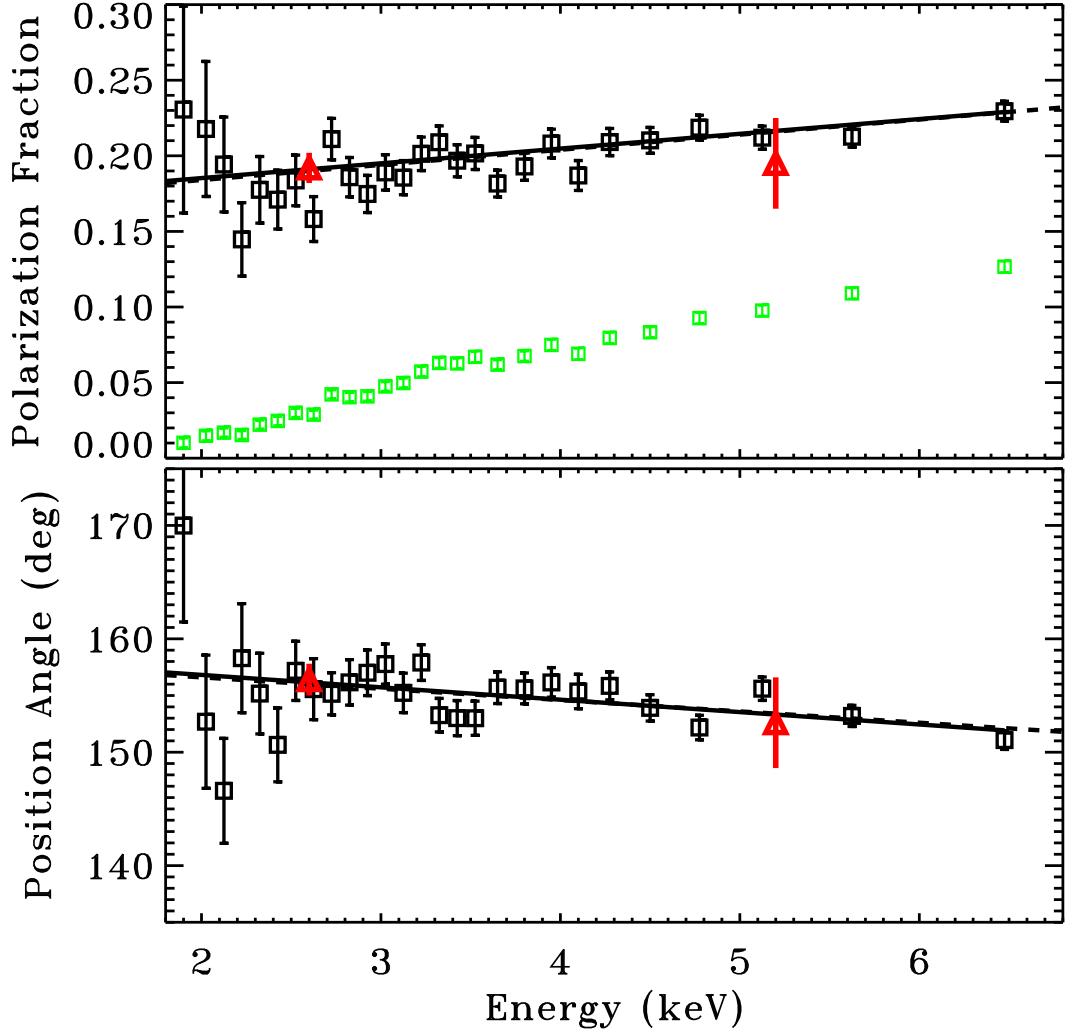}
\caption{Top: Observed fractional modulation amplitude (green symbols)
  and inferred fractional polarization amplitude (black) versus energy
  for our 6 million count Crab nebula simulation described in \S
  5. Here, the observed modulation amplitude (green) is $a(E) =
  (Q(E)^2 + U(E)^2 )^{1/2}/I(E)$, and the inferred polarization
  amplitude is $a_p(E) = a(E)/ \mu(E)$.  The solid (black) line is the
  best fitting polarization amplitude model, and the dashed line shows
  the input amplitude model used to generate the simulation (the
  ``true'' model). Bottom: Observed position angle versus energy for
  our 6 million count Crab nebula simulation. The solid (black) line
  is the best fitting position angle model, and the dashed line shows
  the input model used to generate the simulation (the ``true''
  model). In both panels the red symbols show the existing
  polarization measurements of the Crab nebula as an indication of the
  current state of knowledge (Weisskopf et al. 1978). See \S 5 for
  additional details.  \label{fig1}}
\end{figure}

\newpage

\begin{figure}
\epsscale{0.85}
\plotone{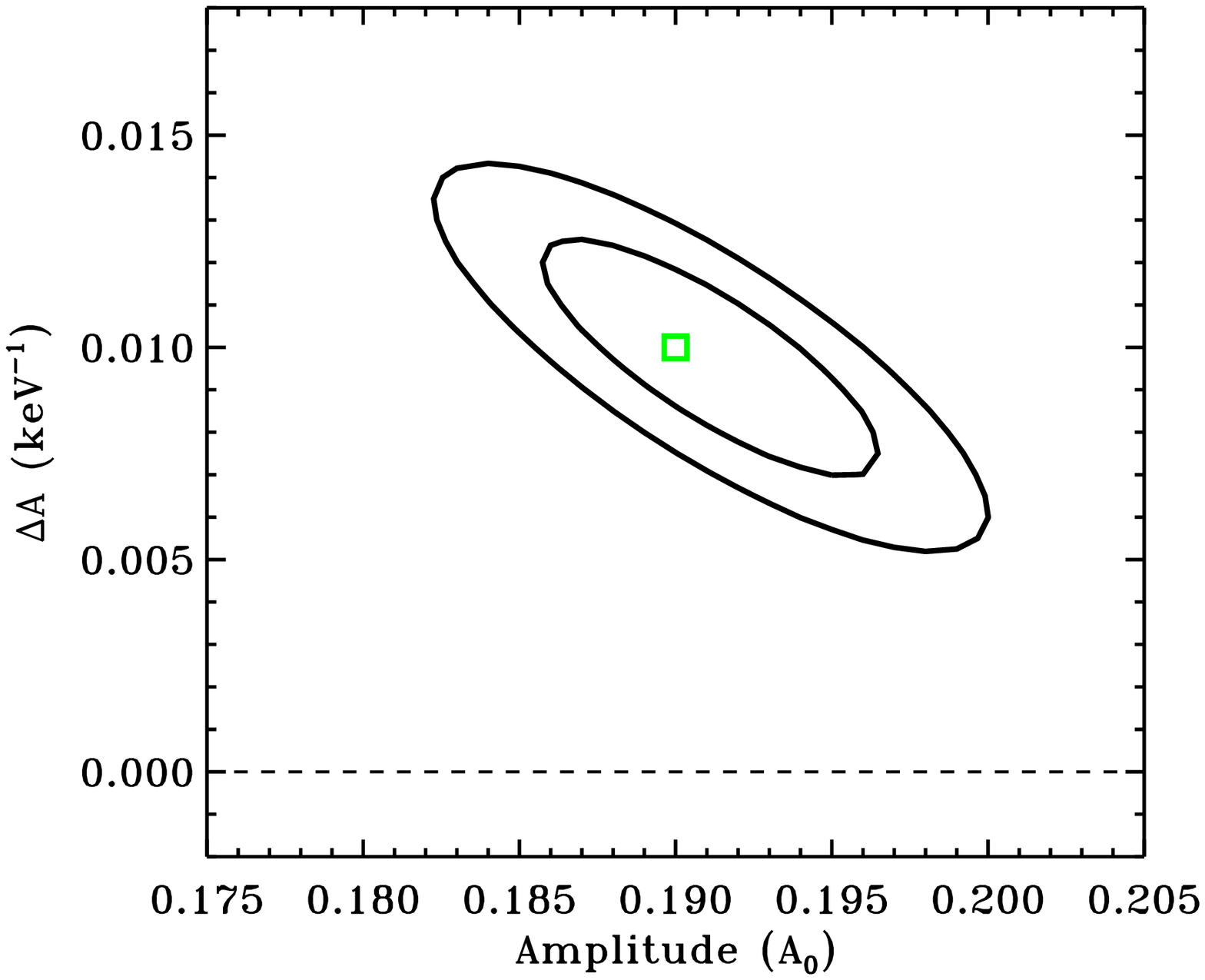}
\caption{Joint confidence regions for the polarization amplitude
  parameters, $A_0$, and $\Delta A$, derived from our 6 million count
  Crab nebula simulation described in \S 5. We show contours drawn at
  $\Delta\chi^2 = 2.3$, and 4.61 which correspond to confidence levels
  of $68.3$, and $90 \%$, respectively. The green square symbol marks
  the input (``true'') values. See the discussion in \S 5 for
  additional details.  \label{fig1}}
\end{figure}

\newpage

\begin{figure}
\epsscale{0.85}
\plotone{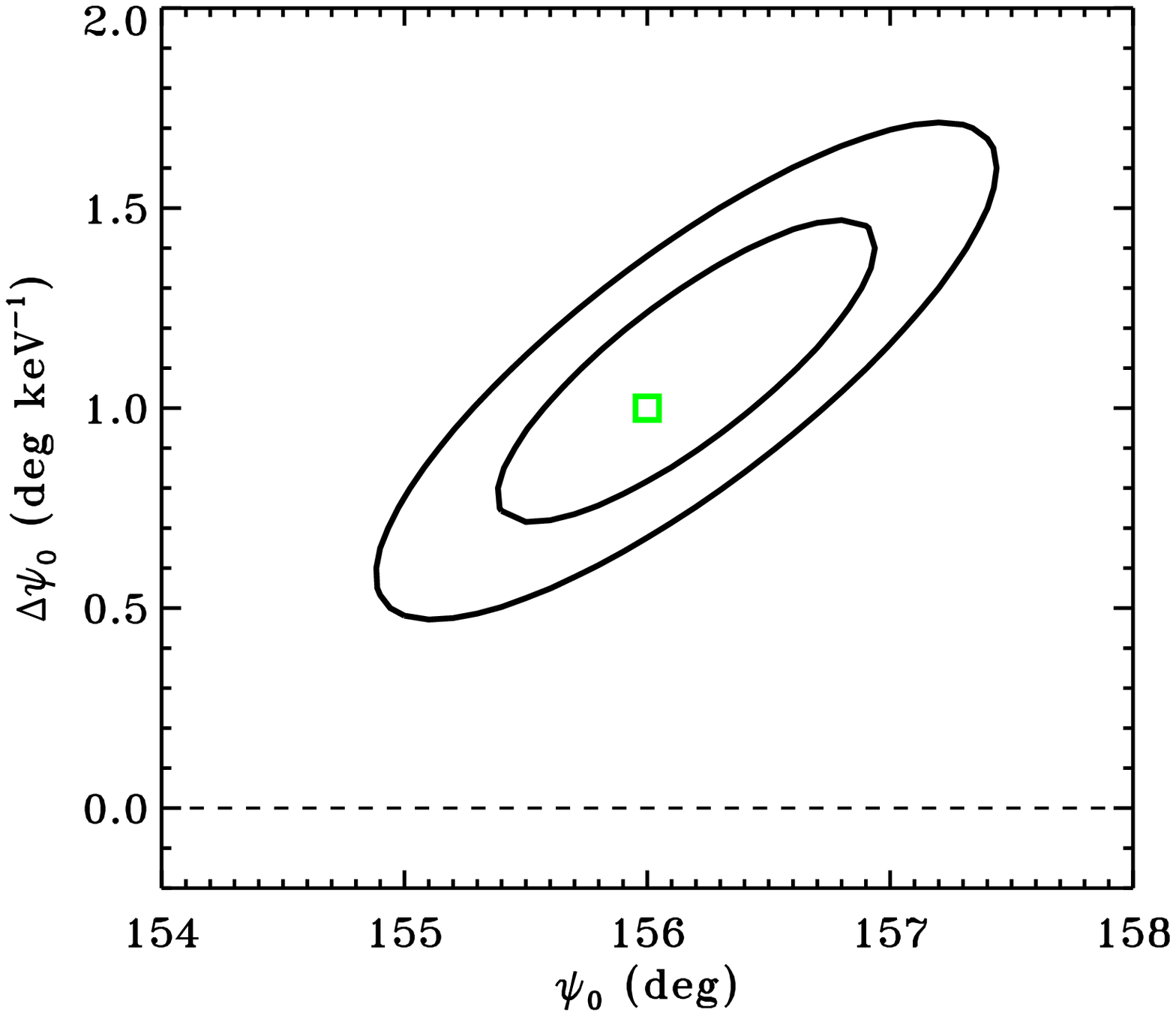}
\caption{Joint confidence regions for the polarization position angle
  parameters $\psi_0$, and $\Delta\psi_0$ for our 6 million count Crab
  nebula simulation described in \S 5. We show contours drawn at
  $\Delta\chi^2 = 2.3$, and 4.61 which correspond to confidence levels
  of $68.3$, and $90 \%$, respectively. The green square symbol marks
  the input (``true'') values. See the discussion in \S 5 for
  additional details.  \label{fig1}}
\end{figure}

\newpage


\end{document}